\documentclass[fleqn,10pt,twocolumn]{wlscirep} %
\usepackage[utf8]{inputenc}
\usepackage[T1]{fontenc}

\usepackage{subfigure} 
\usepackage[abs]{overpic}
\usepackage{nicefrac}
\usepackage{mathtools, cuted} 
\usepackage{multirow}

\usepackage{placeins} 
\usepackage{capt-of} 

\newcommand{\beginsupplement}{%
	\setcounter{table}{0}
	\renewcommand{\thetable}{S\arabic{table}}%
	\setcounter{figure}{0}
	\renewcommand{\thefigure}{S\arabic{figure}}%
}

\newcommand{\SIrefMaterialanalysis}{1}

\title{Real time observation of oxygen diffusion in CGO thin films using spatially resolved Isotope Exchange Raman Spectroscopy}
\author[1,*]{Alexander Stangl}
\author[2]{Nicolas Nuns}
\author[2]{Caroline Pirovano}
\author[3]{Kosova Kreka}
\author[3]{Francesco Chiabrera}
\author[3,4]{Albert Tarancón}
\author[1,*]{Mónica Burriel}
\affil[1]{Univ. Grenoble Alpes, CNRS, Grenoble-INP, LMGP, 38000 Grenoble France}
\affil[2]{Univ. Lille, CNRS, Centrale Lille, Univ. Artois, UMR 8181 – UCCS – Unité de Catalyse et Chimie du Solide, F-59000 Lille, France}
\affil[3]{Catalonia Institute for Energy Research (IREC), Barcelona, Spain}
\affil[4]{ICREA, 23 Passeig Lluis Companys, 08010 Barcelona, Spain}
\affil[*]{alexander.stangl@grenoble-inp.fr, monica.burriel@grenoble-inp.fr}

\begin{abstract}
	
	The exploitation of advanced materials for novel energy, health and computing applications requires deep insight and fundamental understanding of enabling physicochemical mechanisms, such as ionic and electronic conductivity, defect formation processes and reaction kinetics. Therefore, access to the underlying constants of the functional materials via advanced but accessible and straightforward experimental techniques is key.
	Here, we present a novel, cheap, fast and widely applicable approach to analyze oxygen tracer diffusion in thin films with unprecedented time resolution based on the novel in situ isotope exchange Raman spectroscopy (IERS) methodology.
	IERS utilizes the sensitivity of micro Raman spectroscopy to changes in the local isotopic composition, manifested by a frequency shift of the oxygen Raman modes. 
	In-plane tracer diffusion gradients are established by partially blocking the exchange at the surface followed by an isotope exchange annealing.
	Employing a Raman transparent thin surface capping layer allows to follow the resulting isotope exchange and diffusion processes via consecutive spatial and time resolved in situ Raman line scans.
	These isotopic gradients can be analyzed equivalent to diffusion profiles obtained by conventional techniques, such as time-of-flight secondary ion mass spectrometry (ToF-SIMS), to obtain mass-transport coefficients, but with an additional time-component, not accessible by conventional destructive techniques. 
	Here, we study gadolinium doped ceria (CGO) thin films, capped with thin Si$_3$N$_4$ or Al$_2$O$_3$ blocking layers and trenched using a diamond tip.
	We report diffusion coefficients, $D^*$, within the temperature range of interest for intermediate temperature emerging applications from 300\,°C to 500\,°C and confirm the validity of the measurement procedure and extracted parameters by comparison with FEM simulations and literature results.
\end{abstract}

\DeclareUnicodeCharacter{2212}{-}
\begin{document}
\flushbottom
\maketitle

\thispagestyle{empty}
\begin{figure*}[t]
	\centering
	\begin{overpic}[width=170mm, unit=1mm]
		{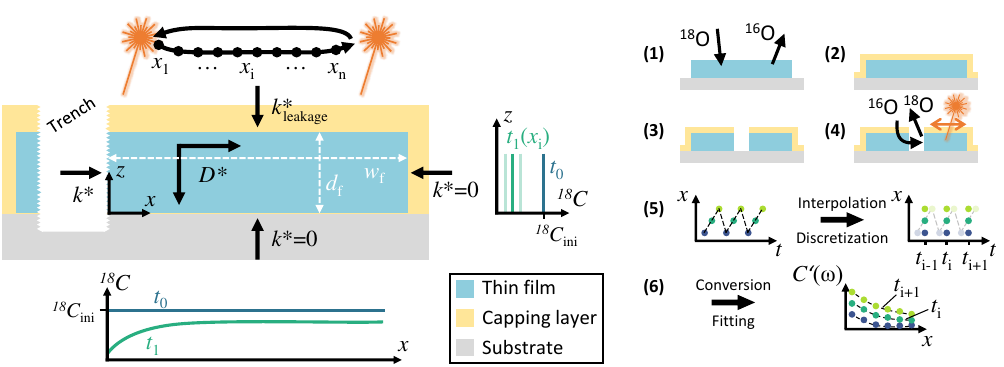}
		\put(0,55){(a)}
		\put(100,55){(b)}
	\end{overpic}
	\caption[]{(a) Schematic of IERS methodology to study oxygen diffusion in thin films. (b) Measurement and data treatment procedure to obtain in-plane diffusion profiles in thin films by IERS: (1) isotope exchange in $^{18}$O enriched atmosphere (2) low temperature deposition of capping layer followed by (3) mechanical trenching and (4) spatially resolved in situ Raman measurements during the isotopic back exchange. (5) Data interpolation, discretization and conversion gives tracer profiles, which can be analyzed within the framework of Fick's diffusion equation.}
	\label{fig:IERS_schematic}
\end{figure*}
\section{Introduction}
Ionic transport processes at the nano- and microscale play a key role in various emerging fields \cite{Maier2005}, such as energy storage and conversion devices, neuromorphic computing, gas sensors and membranes and have attained significant attention among diverse disciplines in science and industry, while meaningful, cheap and widely available analysis techniques are still scarce, especially under in situ and operando conditions \cite{Stangl2021b}. 
One of the most extensively used approaches to study diffusion phenomena is based on the replacement of a chemical element with an isotope of low natural abundance (such as $^{18}$O or $^{6}$Li, so called tracer isotopes) and analyze the tracer distribution within the sample via time-of-flight secondary ion mass spectrometry (ToF-SIMS) \cite{Kilner2011}.
ToF-SIMS is highly sensitive and accurate and commonly used to perform depth profiling to study out-of-plane transport properties in thin films \cite{Navickas2017,Kim2006a,Chen2015a} or along different directions in single crystal \cite{Bassat2013} or polycrystalline or polycrstalline bulk samples \cite{Burriel2012}. Its 3D spatial resolution has previously also enabled the study of in-plane tracer gradients in thin films \cite{Burriel2008b}, wherefore the top surface was blocked utilizing metallic capping layers and subsequent isotope exchange was limited to a lateral surface reopened by trenching. Post exchange analysis allowed to obtain in-plane diffusion and surface exchange coefficients for the first time in epitaxial thin films. 
However, ToF-SIMS is costly, time consuming and destructive and thus limiting its general availability and the reuse of specimens.

To overcome these drawbacks, we take advantage of the non-destructive and cost-effective nature of the recently introduced isotope exchange Raman spectroscopy (IERS) methodology, which is a novel tool to study physicochemical processes in real time \cite{Stangl2023}.
IERS utilizes the sensitivity of micro Raman spectroscopy to study changes in the local isotopic composition, manifested by a frequency shift of the oxygen Raman modes \cite{Guerain2015,Kim1994,Stender2013,Mestl1994}. Raman spectroscopy thus can be used to map the surface of a thin film sample to investigate the in-plane isotopic distribution in a fast and non-destructive manner. 
Compared to conventional techniques, IERS is performed in situ, which gives access to time resolved measurements of the diffusion coefficient with the potential to evaluate the time evolution of transport properties under operation conditions, which can help to identify possible degradation mechanism and enhance material durability.
To establish an in-plane oxygen tracer gradient, thin film samples are coated with a Raman transparent thin capping layer, limiting surface exchange to a lateral surface opened via trenching.
In this work, we study gadolinium doped ceria (CGO) thin films capped with thin Si$_3$N$_4$ or Al$_2$O$_3$ blocking layers and map tracer oxygen in-plane gradients in situ using IERS. We report self-diffusion coefficients and analyze the robustness of the outlined methodology to obtain meaningful results.

\section{Methods}
\subsection{Isotopic sensitivity of Raman spectroscopy and Isotope Exchange Raman Spectroscopy}
Raman spectroscopy is based on the inelastic scattering of monochromatic light and used to investigate the vibrational modes of a molecule or crystal. 
The Raman wavenumber, $\omega$, of the vibration of two crystal atoms, depends within the harmonic approximation on their reduced mass, $\mu$, via $\omega \propto \mu^{\nicefrac{-1}{2}}$ \cite{Colthup1975}. 
Upon isotopic changes in the oxygen sublattice, a continuous frequency shift of the oxygen modes is observed rather than a linear combination of two or three modes for oxygen-cation ($^{16}$O-X,$^{18}$O-X) or oxygen-oxygen ($^{16}$O-$^{16}$O, $^{18}$O-$^{18}$O, $^{16}$O-$^{18}$O) vibrations, respectively.
This single mode behavior is understood within the virtual crystal approximation (VCA), where the masses of the constituting isotopes can be replaced by their weighted average mass, with respect to their relative abundances  \cite{Cardona2005}. The isotopic fraction, ${^{18}C}$, can therefore be determined by measuring the Raman frequency, $\omega ({^{18}\text{O}})$. For the case of a diatomic vibration only involving oxygen \cite{Kim1997}, ${^{18}C}$ can be expressed as:
\begin{equation}\label{eqn:Intro:VCA}
	{^{18}C} = \frac{[^{18}\text{O}]}{[^{18}\text{O}]+[^{16}\text{O}]} = \frac{\left(\frac{\omega_\text{ref}}{\omega ({^{18}\text{O}})}\right)^2-1}{\left(\frac{m_{^{18}\text{O}}}{m_{^{16}\text{O}}}\right)-1},	
\end{equation}
with the wavenumber of the oxygen mode, $\omega_\text{ref}$, corresponding to natural abundance and the atomic weights $m_{^{18}\text{O}}=17.999$\,u and $m_{^{16}\text{O}}=15.995$\,u of $^{18}$O and $^{16}$O, respectively.

IERS utilizes the mass sensitivity of Raman spectroscopy to directly follow changes in the isotopic composition of a sample in situ with time and spatial resolution \cite{Stangl2023}. Therefore, the specimen is annealed in a temperature cell under the microscope and brought in contact with a media of different isotopic concentration (e.g. via the atmosphere) under isothermal conditions to activate isotopic exchange.
For example for surface limited exchange reactions, such as in thin films or (nano-)powders, the analysis of the time evolution of the frequency shift is equivalent to transients obtained by other in situ probes such as electrical conductivity relaxation (ECR) \cite{Chen2003,Cayado2017a}. 
However, the spatial resolution of IERS allows additionally to monitor in-plane tracer gradients for diffusion limited processes in thin films and bulk to extract diffusion coefficients, as outlined in the following section, not achievable by means of any other direct in situ technique.

\subsection{Solution of the 2D cross section diffusion in thin films}
For the study of the in-plane diffusion coefficient, we need to establish a tracer gradient within the thin film. Therefore, a Raman transparent, inert and oxygen exchange blocking layer is conformally deposited on top of the sample and oxygen exchange is only permitted laterally along one well defined line by cutting a trench through the capping layer and the thin film before performing an isotopic (back-)exchange annealing\cite{footnote1}. 
The sample cross section is depicted in Figure~\ref{fig:IERS_schematic}(a), with the thickness, $d_\textrm{f}$ and the sample width, $w_\textrm{f}$. 
$x=0$ is through-out this work defined at the interface of the trench with the thin film, while $z=0$ corresponds to the interface with the substrate.

The in-plane tracer concentration gradients are analyzed in the frame work of Fick's diffusion equations. We assume a homogeneous distribution along the direction of the trench ($y$) which reduces the diffusion model to the 2D cross section along $x$ and $z$: 
\begin{equation}\label{eqn:2D_Fick}
	\frac{\partial^2 c}{\partial x^2}+\frac{\partial^2 c}{\partial z^2}=\frac{1}{D^*}\frac{\partial c}{\partial t}
\end{equation}
with the oxygen tracer concentration, $c={^{18}c}(x,z,t)$, and the tracer self-diffusion coefficient $D^*$. 
In the following, we consider oxygen exchange along the trenched out-of-plane surface at $x=0$, and the possibility of a small leakage exchange through the capped in-plane surface. 
Therefore, the boundary conditions are given by:
\begin{align}\label{eqn:BoundaryConditions}
	-D^*\frac{\partial c}{\partial x} &=k^*_\textrm{CGO}\left(c_\infty-c\right), &x=0, \ 0<z<d_\textrm{f} \\
	-D^*\frac{\partial c}{\partial z} &=k^*_\textrm{leak}\left(c_\infty-c\right), &0<x<w_\textrm{f}, \ z=d_\textrm{f} \\
	-D^*\frac{\partial c}{\partial x} &=0, &x=w_\textrm{f}, \ 0<z<d_\textrm{f} \\
	-D^*\frac{\partial c}{\partial z} &=0, &0<x<w_\textrm{f}, \ z=0
\end{align}
with the surface exchange coefficient, $k^*_\textrm{CGO}$ at the open lateral surface at $x=0$, and $k^*_\textrm{leak}$ accounting for the leakage exchange through the capping layer and the initial and equilibrium concentrations, $c_0$ and $c_\infty$, whereas the latter corresponds to the $^{18}$O concentration of the annealing atmosphere.
We neglect any exchange at the interface with the substrate and any contribution from the second lateral side at $x=w_\textrm{f}$. This can be experimentally guaranteed by ensuring that $w_\textrm{f}$ is much larger than the diffusion length (i.e. semi-infinite diffusion problem). However, we will use the more general (and numerically more stable) solution for the plane sheet model, which is equivalent to the solution of the semi-infinite model for diffusion profiles shorter than $w_\textrm{f}$.
In addition, we assume $c_0$ and $c_\infty$ to be homogeneous.

The solution to Equation~\ref{eqn:2D_Fick} can be expressed as the product of the separated solutions for one dimensional diffusion problems in $x$ and $z$ direction, $c(x,z,t)=c_x\times c_z$ \cite{Jaeger1947}.
Due to the low film thickness, $d_\textrm{f}$, and high $D^*$ of CGO, we assume a homogeneous tracer concentration along $z$ ($D^*\gg d_\textrm{f} k^*_\textrm{leak}$, see discussion below and verification in Figure~\ref{fig:SI:18Ohomogeneity}). Thus, the diffusion problem in out-of-plane direction reduces to the surface limited case, 
while $c_x$ is given as the solution to the plane sheet model \cite{Crank1979}. The normalized isotopic fraction ${C'}$, varying between 0 and 1, thus can be written as
\begin{strip}
	\begin{equation}\label{eqn:FickSolution}
		{C'}(x,z,t) = \frac{c(x,z,t) - c_0}{c_\infty-c_0}=   1-\underbrace{ \left[ \sum_{n=1}^{\infty} \frac{2\delta\cos\left(\beta_n\left(w_\textrm{f}-x\right)/w_\textrm{f}\right)} {\left(\beta_n^2 + \delta^2+\delta\right) \cos\beta_n} \exp \left[\frac{-\beta_n^2Dt}{w_\textrm{f}}\right] \right] }_{{C'}_x}
		\times \underbrace{\exp\left[\frac{-tk^*_\textrm{leak}}{d_\textrm{f}}\right]}_{{C'}_z}
	\end{equation}
\end{strip}
with $\beta_n$ being the $n$'th positive root of the transcendental equation $\beta \tan \left(\beta \right) = w_\textrm{f}k^*_\textrm{CGO}/D^*$ and $\delta={d_\textrm{f}k^*}/{D^*}$.
As can be seen in Equation~\ref{eqn:FickSolution}, ${C'}_z$ reduces to unity for a perfect capping layer with $k^*_\textrm{leak}=0$.

\subsection{Monitoring of the lateral oxygen diffusion using IERS}
A graphical description of the sample preparation steps, IERS measurement procedure and data analysis is given in Figure~\ref{fig:IERS_schematic}(b). 
Small pieces of a thin film sample ($\approx2$\,mm side length) are annealed in an $^{18}$O enriched atmosphere to reach a high homogeneous initial tracer concentration \textbf{(1)}. 
Next, the samples are capped with a Raman transparent, inert and conformal oxygen exchange blocking layer \textbf{(2)}. Low temperature deposition techniques are favored to limit the loss of tracer oxygen during this step. Here, Al$_2$O$_3$ and Si$_3$N$_4$ coatings were employed using atomic layer deposition (ALD) and plasma enhanced chemical vapor deposition (PECVD), respectively, with deposition temperatures between 200 and 280\,°C. 
A trench of approximately 3\,µm is cut through the capping and thin film layer using a precision diamond scriber in order to enable oxygen exchange at the lateral side of the CGO thin film \textbf{(3)}. Mechanical scratching was selected to avoid thermal or chemical modifications of the CGO surface, as may occur using focused ion beam or chemical etching techniques.

The following back-exchange in dry air is monitored using continuous Raman line scans perpendicular to the direction of the trench \textbf{(4)} using a motorized linear stage. The movement of the oxygen tracer becomes directly visible via the time and spatial dependent frequency shift of the oxygen Raman mode, which is obtained via line fitting.

The line scan parameters (scanning length and point spacing) have to be adopted to expected $D^*$ and $k^*$ parameters, for example via prior finite element method (FEM) simulations using literature values. 
To improve time resolution, a variable point spacing has been used, e.g. increasing spacing with increasing distance from the trench. Raman acquisition times have to be reasonably short, while laser induced heating has to be avoided by restricting the laser power to a material dependent maximum.

At small times, $t$ and fast kinetics, there may be large changes in $\omega$ between two spectra measured at the same location (e.g. after one completed line scan) and changes between the first spectrum and the last in one single line scan may be significant. 
To obtain meaningful diffusion profiles at the \textit{same} discrete time, $t_i$, the transients for each position, $\omega_\textrm{max}(x_i)$, are interpolated using an analytical function and discretized at the same discrete times, $t_\textrm{i}$ (see Figure~\ref{fig:SI:IERS}).
The isotopic fraction, $^{18}C$, and consequently the $^{18}$O normalized isotopic fraction, ${C'}$, are determined using the VCA model and Equation~\ref{eqn:Intro:VCA} \textbf{(6)}. The reference values, $\omega_\textrm{ref}(T)$, are determined by temperature calibration of the frequency shift of a sample with natural $^{18}$O abundance.
The resulting in-plane diffusion profiles at different discrete times are fitted to the solution of the diffusion equation (Equation~\ref{eqn:FickSolution}) to obtain the transport coefficients.

Note, in this work in situ experiments were performed during back-exchange, as this allows to exchange various samples at the same time and thus significantly reduces the consumption of tracer oxygen gas, while the described approach is equally valid to study the initial exchange.

Further details can be found in the Experimental Section at the end of the manuscript.

\section{Results and discussion}
\subsection{Structural characterisation and isotope profiling}
Polycrystalline 120\,nm-thick dense Ce$_{0.8}$Gd$_{0.2}$O$_{2-\delta}$ films were deposited by PLD on platinized Si substrates (see experimental for more details). The obtained films are of low roughness ($R_\text{q} = 1.6$\,nm) with nanometric grains of about $48\pm23$\,nm. An overview of their structural and chemical characteristics is shown in \textbf{Figure~S\SIrefMaterialanalysis}.
The cubic fluorite structure (O$_\text{h}^5$ (Fm-3m) space group) of cerium oxide leads to one Raman active phonon mode of T$_\text{2g}$ symmetry. It results from the symmetric stretching vibrations of the cation-oxygen O-Ce-O units and therefore only involves the motion of oxygen atoms. 
A representative Raman spectrum of an as deposited CGO thin film with natural $^{18}$O isotopic abundance (0.2\,\%) is shown in Figure~S\SIrefMaterialanalysis(a). The wavenumber of the main mode (I) is about 465\,cm$^{-1}$, as observed for undoped ceria. The long tail on the left shoulder of the main band and additionally observed modes at slightly higher frequencies (II \& III) are caused by the Gd doping, which reduces the symmetry of the crystal lattice \cite{McBride1994}. 
The metallic Pt buffer layer blocks any signal coming from the Si substrate.

\begin{figure*}[]
	\centering
	\begin{minipage}[b]{0.49\textwidth}
		\begin{overpic}[width=\textwidth, unit=1mm]
			{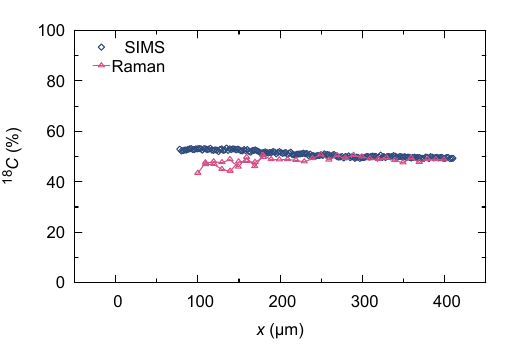}
			\put(-2.5,55){(a)}
		\end{overpic}
	\end{minipage}
	\begin{minipage}[b]{0.49\textwidth}
		\begin{overpic}[width=\textwidth, unit=1mm]
			{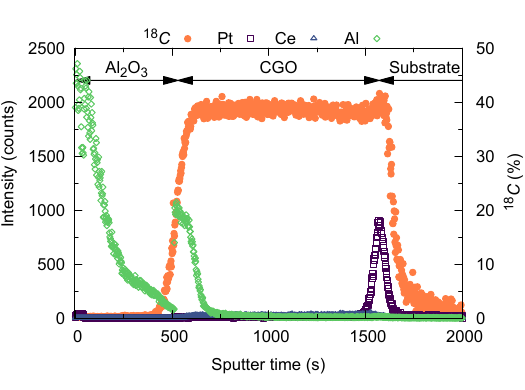}
			\put(-2.5,55){(b)}
		\end{overpic}
	\end{minipage}
	\caption[]{(a) In-plane distribution of the isotopic fraction of $^{18}$O in a CGO thin film after the exchange annealing and the deposition of an Al$_2$O$_3$ capping layer by ALD obtained via SIMS and Raman mapping (integrated along $y$, arbitrary $x$ position) and (b) corresponding out-of-plane ToF-SIMS ion depth profiles, showing a homogeneous and isotropic distribution of the tracer along both directions before trenching and performing the back-exchange.
	}
	\label{fig:SI:18Ohomogeneity}
\end{figure*}

\begin{figure*}[t]
	\centering
	\begin{overpic}[width=89mm, unit=1mm]
		{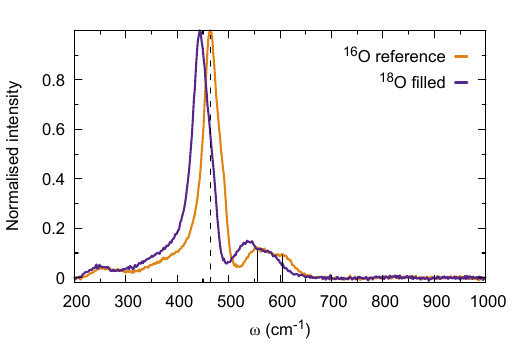}
	\end{overpic}
	\caption[]{$^{18}$O induced Raman frequency shift of the CGO modes. The $^{16}$O reference corresponds to an as deposited sample with natural $^{18}$O abundance.}
	\label{fig:SI:Ramanshift}
\end{figure*}

The CGO films were filled with oxygen tracer by annealing in $^{18}$O enriched atmosphere and subsequently capped with an oxygen exchange blocking layer. Using low temperature deposition techniques minimized changes in $^{18}C$ during the capping process as verified via Raman measurements before and after the deposition.
ToF-SIMS analysis revealed a homogeneous and isotropic tracer distribution in-plane as well as out-of-plane, see Figure~\ref{fig:SI:18Ohomogeneity}(a \& b). The flat $^{18}C$ depth profile indicates that the overall exchange process is limited by a surface reaction (in the uncapped sample) and out-of-plane diffusion is comparably fast. This, together with a homogeneous, initial in-plane distribution, is a requirement for the further analysis.
The isotopic induced red shift of the T$_\text{2g}$ peak and the two defect modes is shown in Figure~\ref{fig:SI:Ramanshift}, which confirms that all three modes only involve the motion of oxygen atoms.

\begin{figure*}[t]
	\centering
	\begin{overpic}[width=89mm, unit=1mm]
		{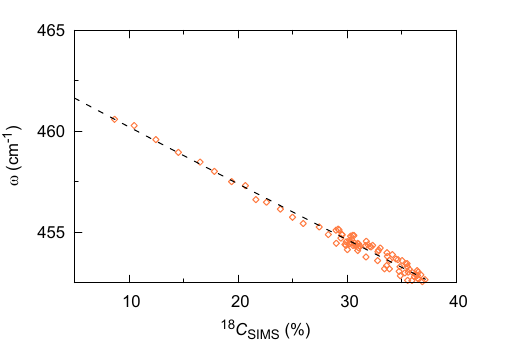}
		\put(0,0){\includegraphics[width=89mm]{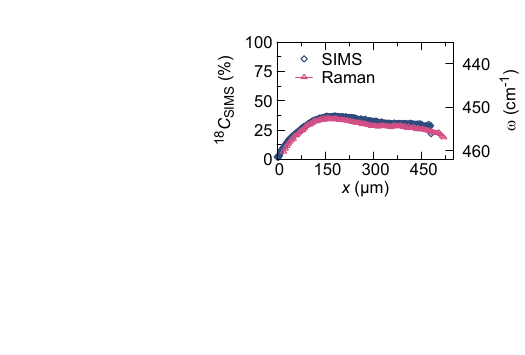}}
	\end{overpic}
	\caption[]{Raman frequency shift as a function of the $^{18}$O isotopic fraction. The dashed line corresponds to the best fit to the virtual crystal approximation. This continuous calibration curve is obtained by mapping the same sample area with an $^{18}C$ in-plane profile using ToF-SIMS and Raman spectroscopy, as shown in the inset, and correlating the two measurements via the $x$-axis. Note, that shown data for both techniques is integrated along the $y$-axis. $x=0$ corresponds to the edge of the trench.}
	\label{fig:confirmation}
\end{figure*}
To show the continuous shift of the Raman mode with the isotopic fraction, we have mapped the same area of a sample with an in-plane isotopic gradient using Raman spectroscopy and ToF-SIMS. The inset of Figure~\ref{fig:confirmation} shows the $x$ distribution of $^{18}C$ as obtained by SIMS measurement, as well as the corresponding Raman mode position. Both datasets were averaged along $y$ to reduce the error arising from the uncertainty in selecting the same sample area. Combining both measurements for each $x$ position, as displayed in the main panel, shows a very good correlation between the frequency shift and the isotopic fraction, as expected within the VCA. Fitting this data to Equation~\ref{eqn:Intro:VCA} (dashed line) gives ${\omega_\text{ref}}=463$\,cm$^{-1}$, which serves as room temperature calibration reference to directly assess the isotopic concentration, $^{18}C$, from Raman measurements with high accuracy.

\subsection{Impact of leakages in the blocking layer}
\begin{figure*}[!hb]
	\centering
	\begin{overpic}[width=170mm, unit=1mm]
		{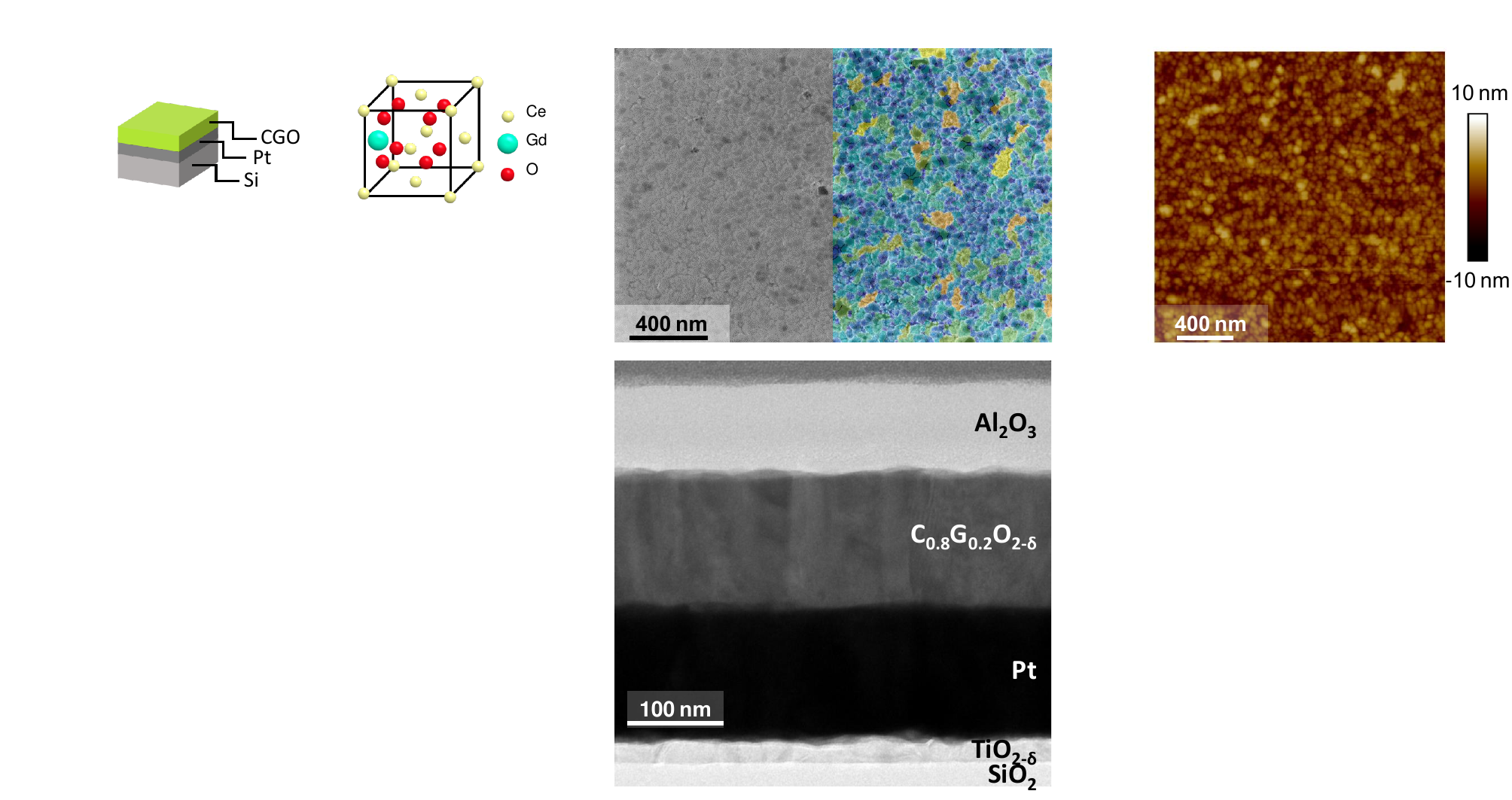}
		\put(0,0){\includegraphics[width=170mm]{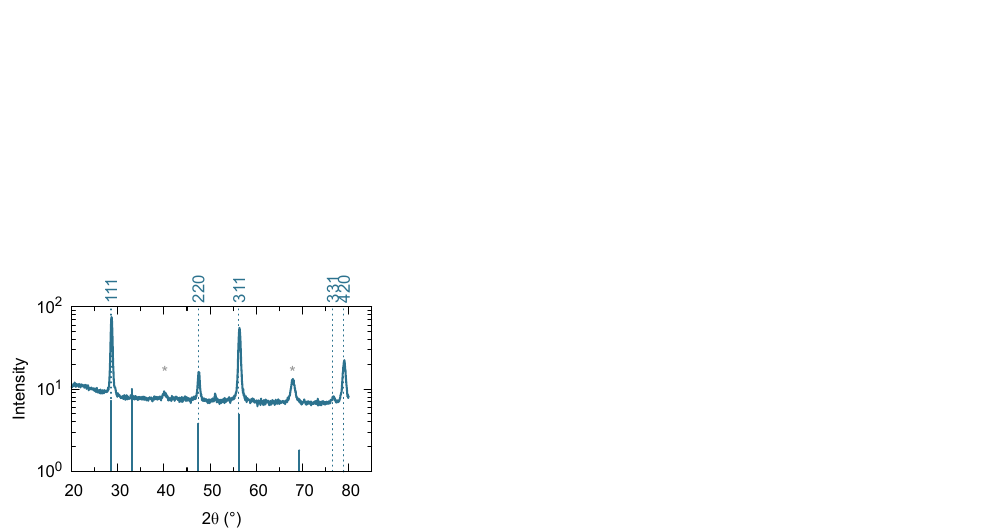}}
		\put(0,0){\includegraphics[width=170mm]{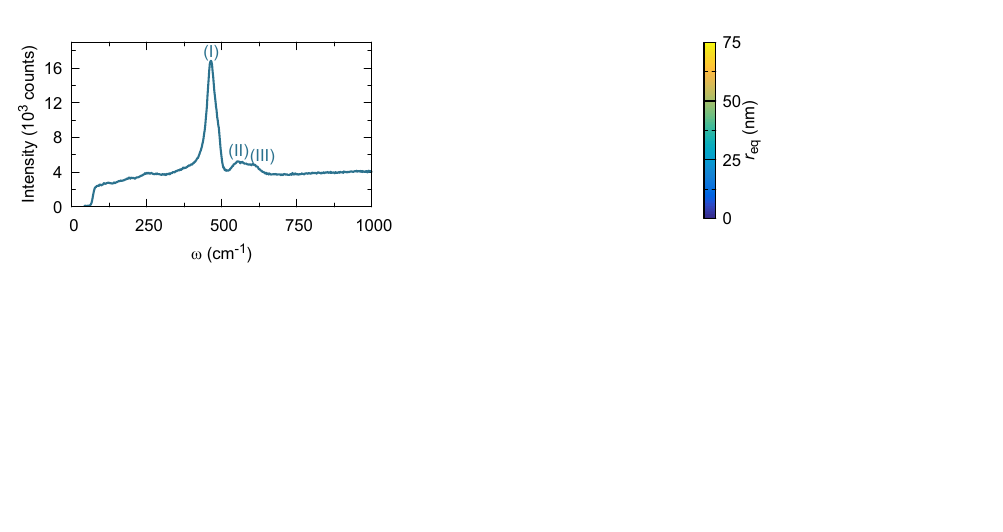}}
		\put(0,0){\includegraphics[width=170mm]{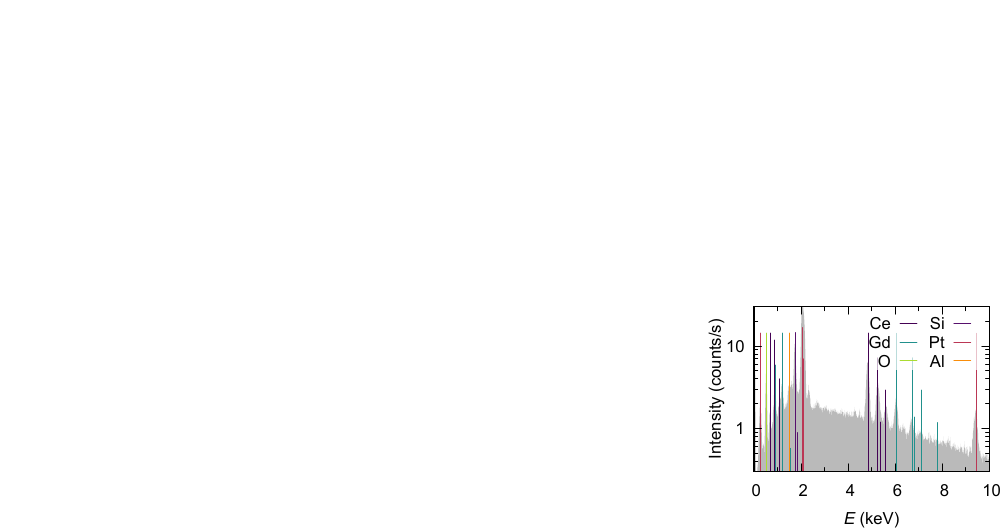}}
		\put(2,86){(a)}
		\put(2,44){(b)}
		\put(63,86){(c)}
		\put(63,44){(d)}
		\put(127,86){(e)}
		\put(119,44){(f)}
	\end{overpic}
	\caption[]{\textbf{Material characterisation of the CGO/Pt/Si thin films:} (a) Raman spectrum of CGO composed of the oxygen breathing mode (I) and defect modes (II \& III) induced via Gd substitution. Inset shows schematic of sample architecture and CGO crystal structure. (b) Grazing incidence X-ray diffraction pattern showing textured growth. Films are free of any secondary phases. Substrate peaks at 40.2 and 67.9\,° are marked with asterisks. (c) Secondary electron SEM top view image reveals a homogenous, dense structure, with nanometric grain sizes of $48\pm23$\,nm. False colour plot shows equivalent grain radius, $r_\text{eq}$. (d) Cross-section transmission electron microscopy (TEM) image of 120\,nm CGO film on top of Pt with 90\,nm alumina coating. (e) Topographical atomic force microscopy (AFM) surface map. (f) Energy dispersive X-ray spectrum of a CGO sample.}
	\label{fig:SI_Material_analysis}
\end{figure*}

We have tested Al$_2$O$_3$ and Si$_3$N$_4$ thin films, deposited by ALD and PECVD, respectively, as blocking layer on top of $^{18}$O filled CGO samples. The layer thicknesses were varied between 20 and 100\,nm. Higher thicknesses were avoided to minimize any detrimental influence on the Raman signal, such as increased background and decreased signal-to-noise ratio. A cross sectional TEM image of the sample architecture is shown in Figure~\ref{fig:SI_Material_analysis}(d) and a SEM/EDX top view analysis of the trench cut into the sample is shown in Figure~\ref{fig:SI:SEM_trench}, confirming that the trench cuts through the full multi-layer structure and reaches the Si substrate. 

\begin{figure*}[]
	\centering
	\begin{minipage}[b]{0.49\textwidth}
		\begin{overpic}[width=\textwidth, unit=1mm]
			{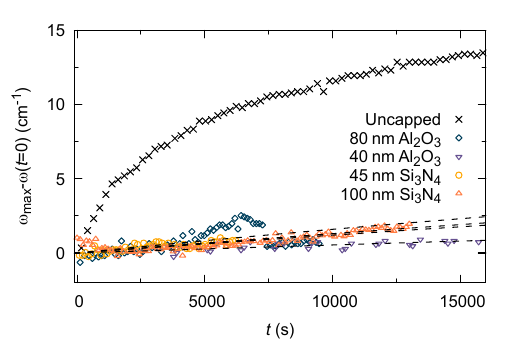}
			\put(-2.5,55){(a)}
		\end{overpic}
	\end{minipage}
	\begin{minipage}[b]{0.49\textwidth}
		\begin{overpic}[width=\textwidth, unit=1mm]
			{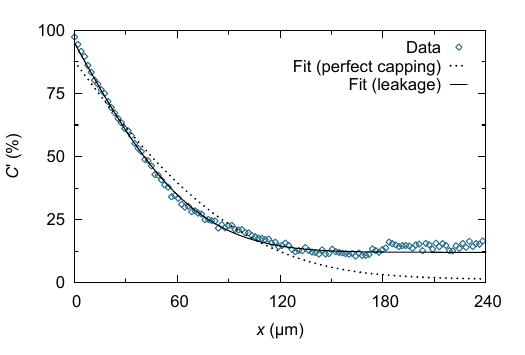}
			\put(-2.5,55){(b)}
		\end{overpic}
	\end{minipage}
	\caption[]{Analysis of leakage through surface coating: (a) in situ IERS measurements to study the time evolution of the T$_\textrm{2g}$ mode at 400\,°C of CGO thin films capped with blocking layers of different materials and thicknesses deposited by ALD or PECVD, before opening a trench through the surface. Small changes in $\omega_\textrm{max}$ point towards small exchange fluxes through the capping layer, \textit{i.e.} a small leakage. Assuming a simple surface limited reaction, $k^*_\textrm{leak}$ can be assessed (dashed lines). Values range from $1\times10^{-11}$ to $10\times10^{-11}$\,cm\,s$^{-1}$. (b) Tof-SIMS diffusion profile after back-exchange at 400\,°C of an Al$_2$O$_3$ capped and trenched CGO thin film along $x$. The open surface (trench) is at $x=0$. Fitting is performed either assuming perfect capping (dotted line) or a small leakage through the surface (solid line), respectively. The introduction of the surface exchange coefficient, $k_\textrm{leak}\approx3\times10^{-11}$, for the leakage through the capping layer in $z$ direction, significantly reduces the fit error and allows an appropriate reproduction of experimental data. Additionally, the derived fitting value matches well the numbers obtained for leakage tests using in situ Raman of various capping layers.}
	\label{fig:SI:leakage_analysis}
\end{figure*}

The layer tightness was tested using in situ IERS measurements at 400\,°C in dry air (natural $^{18}$O abundance), as shown in Figure~\ref{fig:SI:leakage_analysis}(a).
If the capping layer perfectly blocks the exchange with the atmosphere, one would expect no changes in the Raman mode position with time at constant temperature. However, for all tested materials, we find a slow increase of $\omega_\textrm{max}$, which corresponds to the loss of $^{18}$O and thus imperfect capping. Assuming a simple surface limited reaction, one can estimate the leakage rate through the blocking layer (see SI for details). The resulting surface exchange coefficients for the leakage processes, $k^*_\textrm{leak}$, are in the range of $1\cdot10^{-11}$ to $10\cdot10^{-11}$\,cm\,s$^{-1}$, with no significant differences between Al$_2$O$_3$ and Si$_3$N$_4$ deposited by different techniques (the lowest value corresponds to a 40\,nm Al$_2$O$_3$ capping layer by ALD at 200\,°C). In comparison, this is significantly slower than the uncovered CGO surface with $k^*\approx1\cdot10^{-9}$.
Still, this leakage needs to be considered when analyzing in-plane diffusion profiles. 

An exemplary ToF-SIMS tracer gradient after back-exchange at 400\,°C is shown in Figure~\ref{fig:SI:leakage_analysis}(b). $x=0$ corresponds to the trench position. The measurement was cut off after around 240\,µm, while the sample width was approx. 2\,mm. The profile can be separated into two regions. First, for $x<150$\,µm, the curve follows the shape of a classic diffusion profile. For larger distances from the open surface, the profile is flat, but non-zero. This behavior cannot be explained with a simple one-dimensional diffusion process, with oxygen exchange occurring only at the lateral side surface at $x=0$, but is well described via a mechanism taking into account a small leakage through the capped surface, as shown via fitting of the diffusion Equation~\ref{eqn:FickSolution} for the case of leakage with a reduced surface exchange coefficient, $k_\textrm{leak}$ compared to perfect capping with ${C'}_z=1$. The first solution matches well the experimental data along the full profile, with $k_\textrm{leak}\approx3\times10^{-11}$ ($D^*=3\cdot10^{-10}$\,cm$^2$\,s$^{-1}$ and $k^*=8\cdot10^{-7}$\,cm\,s$^{-1}$). This value is similar to the leakage rates observed above by in situ IERS and confirms that the employed capping layers are not fully tight. 
However, it is clearly visible, that the exchange within the top surface is strongly reduced, leading to an $^{18}$O in-plane gradient, which can be successfully fitted to obtain the oxygen transport coefficients.

Note, that we have observed the formation of blisters in the capping layer during the heating to the annealing temperature, which locally enabled faster oxygen exchange. These may have been caused by impurities on the sample surface, which resisted the cleaning procedure before the deposition of the capping layer or by trapped gas during the deposition. 
For details see Figure~\ref{fig:SI:blister_coating} in the supplementary information. As these defects are easily visible by the optical microscope of the Raman setup, they can be avoided when selecting the region for the in situ IERS measurements, as described next.

\subsection{In situ IERS measurements and calculation of the mass transport properties}
\begin{figure*}[!th]
	\centering
	\begin{overpic}[width=170mm, unit=1mm]
		{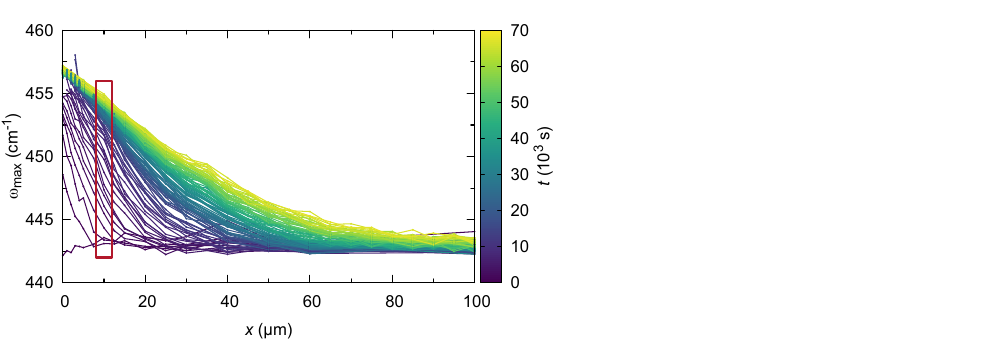}
		\put(0,0){\includegraphics[width=170mm]{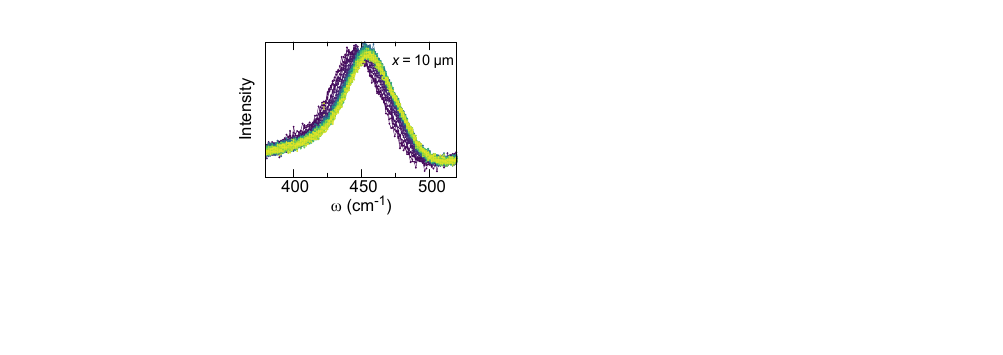}}
		\put(0,0){\includegraphics[width=170mm]{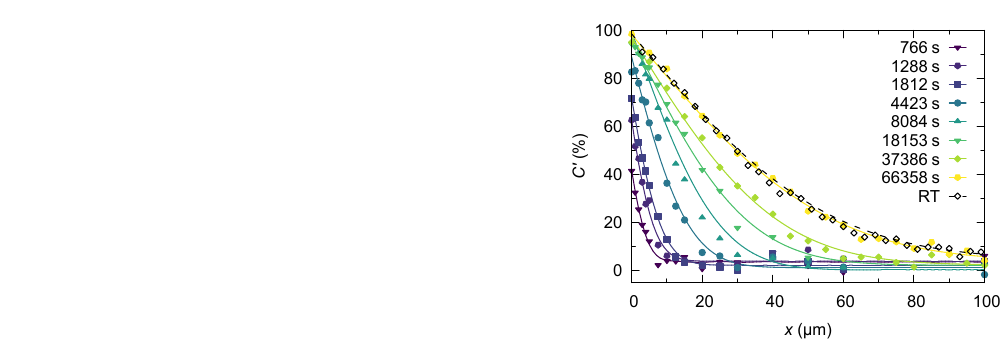}}
		\put(-2,53){(a)}
		\put(95,53){(b)}
	\end{overpic}
	\begin{overpic}[width=170mm, unit=1mm]
		{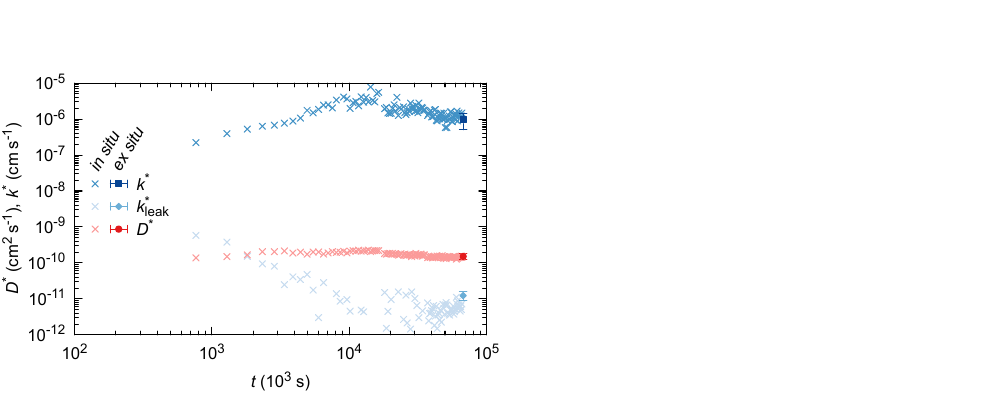}
		\put(0,0){\includegraphics[width=170mm]{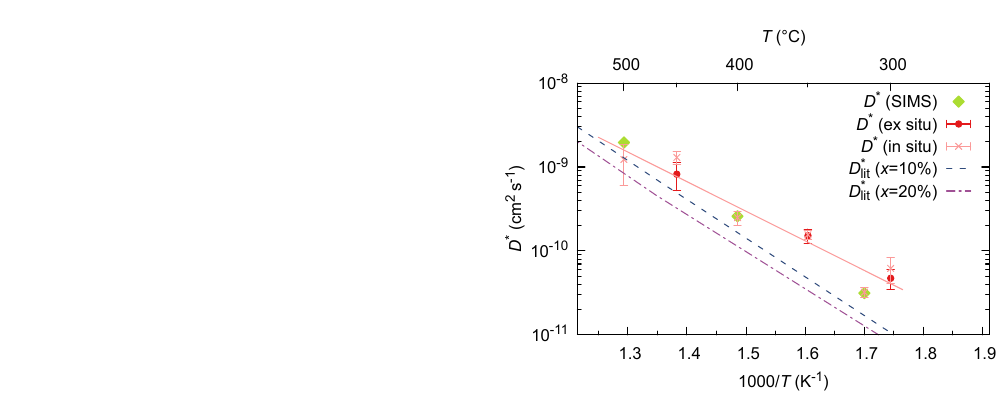}}
		\put(-2,53){(c)}
		\put(85,53){(d)}
	\end{overpic}
	\caption[]{\textbf{In situ in-plane IERS measurements: } (a) Spatial and time resolved Raman frequency shift of T$_\text{2g}$ mode during a back-exchange of trenched SiN/CGO thin film at 350\,°C obtained by repeated line scans. Inset shows the time evolution of the Raman spectrum for $x=10$, marked with a red rectangular. (b) In-plane diffusion profiles of oxygen tracer for various times, determined from (a). Solid lines are best fits to the diffusion equation. Black symbols/line are obtained ex situ at room temperature after the back-exchange. (c) Time evolution of the fitted transport coefficients at 350\,°C from in situ Raman measurements and comparison with ex situ values from room temperature mapping. (d) Temperature dependence of $D^*$ values from SIMS, in and ex situ Raman measurements. Solid line is an Arrhenius fit to the in situ data ($\times$).
		Dashed lines correspond to data found in literature for 10\% \cite{Manning1996} (extrapolated to lower temperatures) and 20\%\cite{Kowalski2009} doped CGO bulk.}
	\label{fig:IERSinplane}
\end{figure*}
IERS measurements were performed as described in the experimental section using a temperature cell mounted to the Raman stage under flowing dry air. 
A spatially and time resolved in situ IERS measurement is shown in Figure~\ref{fig:IERSinplane}(a) for a trenched SiN/CGO sample at 350\,°C during a back-exchange experiment. Each line scan consists of 15 to 25 spectra, with variable spacing (1\,µm close to the trench, 5--20\,µm at the right end), a scanning length of 100\,µm and a total duration of about 10\,min.
The time evolution of the T$_\text{2g}$ mode for one specific distance from the surface ($x=10$\,µm) is depicted in the inset, showing a clear shift to higher wavenumbers with time. The Raman mode frequency, $\omega_\textrm{max}(x_i,t)$ is obtained by line fitting, as shown in the supplementary information (Figure~\ref{fig:SI:IERS}(a)) using a single Voigt profile.
Starting at the opened lateral surface at $x=0$, a diffusion profile starts building up with time as the $^{18}$O concentration reduces in the film due to annealing in an $^{16}$O atmosphere.

The transients are normalized with respect to the initial tracer concentration in the sample before the back-exchange and the saturation value, which corresponds to the $^{16}$O reference position and are then converted to the normalized isotopic fraction ${C'}$, as shown in Figure~\ref{fig:IERSinplane}(b), for some selected discrete times. The solution of the 2D diffusion equation (Equation~\ref{eqn:FickSolution}) is used for fitting the in-plane tracer gradients.
As can be seen, generally experimental data can be well reproduced over the whole time range of the experiment, as shown with solid lines.
At the end of the in situ experiment, the same sample area was mapped at room temperature using Raman spectroscopy and the resulting ${C'}$ profile is shown with black open diamonds (RT). The remarkable resemblance to the final high temperature data validates the in situ approach presented here.

One of the main advantages of any in situ methodology is its time resolution, while post mortem analysis only illuminates a single time frame. In situ IERS allows to study the evolution of transport properties with time, as shown in Figure~\ref{fig:IERSinplane}(c) for $k^*$, $k^*_\textrm{leak}$ and $D^*$ over about 19\,hours. The diffusion coefficient remains constant, while we observe some variations with time for the fitted surface exchange coefficients. Notably, for all three coefficients the final parameters at high temperature match well ex situ values from the RT analysis.

\begin{figure*}[t]
	\centering
	\begin{overpic}[width=89mm, unit=1mm]
		{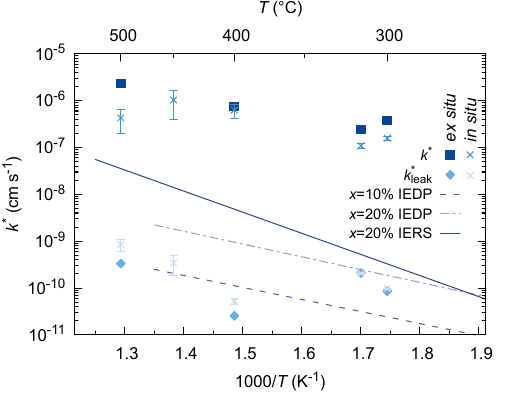}
	\end{overpic}
	\caption[]{Arrhenius plot of $k^*$ coefficients obtained by in-plane mapping: \textit{ex situ} via SIMS and room temperature Raman (solid symbols) and in situ Raman ($\times$). Blue lines correspond to literature: 10\%CGO IEDP (dashed \cite{Manning1996}), 20\%CGO IEDP (double broken \cite{Kowalski2009}) and IERS (solid \cite{Stangl2023}).}
	\label{fig:Si:Arrhenius}
\end{figure*}

The exchange rate through the capping layer is approximately $10^{-11}$\,cm\,s$^{-1}$ at 350\,°C, which is slightly lower than the values found above at 400\,°C and therefore plausible. 
On the other hand, the lateral surface exchange coefficient is very large with about $10^{-6}$\,cm\,s$^{-1}$. Values reported for CGO in literature were found in the range of $10^{-9}$ -- $10^{-8}$\,cm\,s$^{-1}$ around 350\,°C \cite{Stangl2023,Kowalski2009} and thus, more than two orders of magnitude smaller. As shown in the Arrhenius plot of the surface exchange coefficients in Figure~\ref{fig:Si:Arrhenius}, this trend persists within the whole analyzed temperature range between 300 and 500\,°C.
To exclude possible catalytic influences of the Pt buffer layer, we have reproduced the experiment with a twin CGO thin film sample grown on a MgO single crystal substrate with similar results.
As the samples are polycrystalline, intrinsic differences between the lateral and the top surfaces are not expected, however, the fact that the trench was opened just before the back-exchange may play a possible role. The detrimental influence of surface contamination on exchange rates is well documented in literature and pristine surfaces were found to be significantly more active than their aged counterparts due to the exposure to different atmospheres, annealing conditions or contaminants \cite{Acosta2022,Rupp2017,Siebenhofer2020a}. We have observed previously, that alumina deposited by ALD is strongly reduced and may therefore lower the total oxygen content of the CGO layer as well, which would result in an additional chemical driving force, with a chemical surface exchange coefficient, $k^\delta$, and thus lead to faster overall kinetics. However, this is not expected for Si$_3$N$_4$, while the obtained exchange coefficients do not differ significantly between the different capping materials.

The triple phase boundary between the capping layer, the CGO thin film and the atmosphere at the trench may enhance the exchange rate but this remains beyond the scope of this work. 
Additionally, as will be discussed in the next section, the extraction of the lateral surface exchange coefficient based on the described approach is error-prone due to uncertainties in the definition of the origin and possible chipping off of the coating close to the trench due to the mechanical trenching process and bad adhesion.

But first, we focus on the analysis of the diffusion coefficient. At 350\,°C $D^*$ is approximately $1\cdot10^{-10}$\,cm$^2$\,s$^{-1}$, which is close to literature values for bulk CGO. A comparison for in situ and ex situ measurements and literature over the full measured temperature range is shown in Figure~\ref{fig:IERSinplane}(d). We find thermally activated behavior of $D^*$ and striking agreement between in situ line scans and ex situ values obtained by either Raman or SIMS mapping. The solid line is an Arrhenius fit to the in situ data with an activation energy of approximately 0.7\,eV, compared to 0.9 -- 1.2\,eV found in literature \cite{Manning1996,Kowalski2009,Ruiz-Trejo1998}. These small differences are likely originated in the different material types studied (single crystal and ceramics vs. thin film) and small variations in the Gd doping, which are known to impact the process of oxygen transport. 

\subsection{Factors affecting the determination of the mass transport properties}

\begin{figure*}[!th]
	\centering
	\begin{minipage}{85mm}
		\begin{overpic}[width=89mm]
			{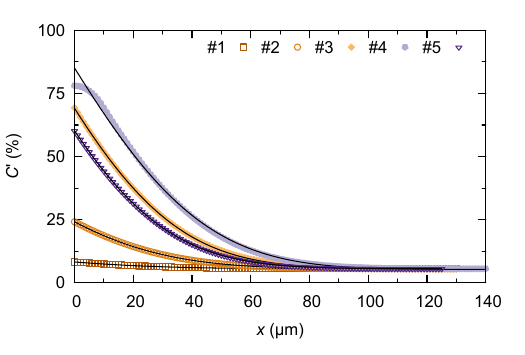}
			\put(0,0){\includegraphics[width=89mm]{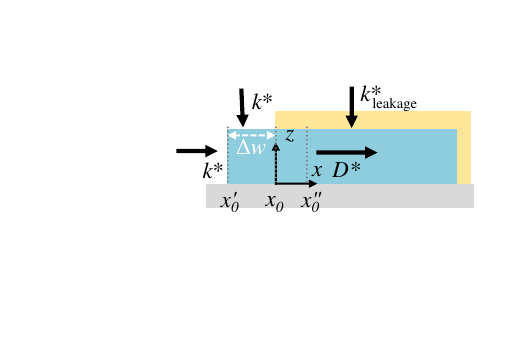}}
		\end{overpic}
	\end{minipage}
	\begin{minipage}{90mm}
		\begin{overpic}[width=\textwidth, unit=1mm]
			{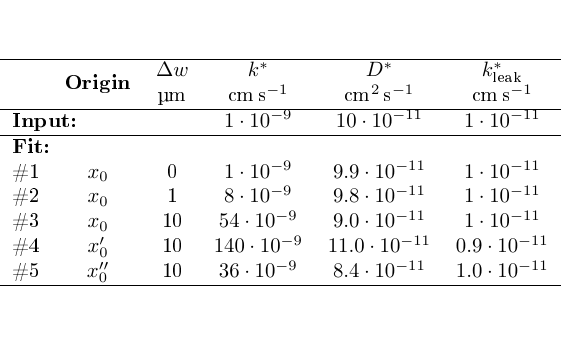}
		\end{overpic}
	\end{minipage}
	\caption[]{Finite element model simulations to analyze the influence of a partially uncovered top surface and the resulting uncertainty in defining the origin for five different scenarios varying the width of the uncovered top surface $\Delta w$ between 0 and 10\,µm and the origin for the fit. Solid lines are fits of the diffusion equation to the simulated data. 
		Input and fitted values for $k^*$, $D^*$ and $k_\textrm{leak}^*$ are given in the Table on the right ($\Delta w$ is varied for the different simulations, while the origin is shifted for the fittings \#4 and \#5). The simulated annealing time is 67800\,s.}
	\label{fig:COMSOL}
\end{figure*}

We found very good correlation between SIMS, in situ and ex situ Raman measurements and literature values for the diffusion coefficient, $D^*$, and matching values for $k_\textrm{leak}$ for different types of experiments. However, the lateral surface exchange coefficient of CGO, $k^*$, was significantly larger than expected. 
We have identified two potential factors which may affect the accuracy of the obtained transport properties. First, the blocking layer may be partially removed due to chipping off during the mechanical trenching. And second, uncertainties in defining the origin, i.e. the position of the trench, may arise due to small movements of the sample during the measurement due to thermal expansion of the sample and the heater or precision errors in the repeated positioning of the stage. While movements towards the trench are easily detected by the loss of the Raman signal of the material and the appearance of the substrate peak, shifts towards the sample are not trivially identified.

To analyze the influence of these issues and verify the robustness of the method and the validity of the obtained results, we have performed FEM simulations using COMSOL. A schematic of the employed 2D model is shown in the inset of Figure~\ref{fig:COMSOL}, which approximates the real sample geometry. The width of the unblocked surface area is given by $\Delta w$ and was varied between 0 and 10\,µm, which corresponds to about 3$\times$ the trench width.
We have tested various combinations for the input parameters $k^*$, $D^*$ and $k_\textrm{leak}^*$, matching experimental results at different temperatures. The input values for the simulated data displayed in Figure~\ref{fig:COMSOL} correspond to about 350\,°C and are given in the Table of Figure~\ref{fig:COMSOL}.
Note that we used the same exchange coefficient, $k^*$ for the unblocked top and the opened lateral surface for the simulation, as the studied thin films are of polycrystalline structure. 

Upon partially uncovering the surface, the isotopic fraction inside the thin film drastically increases (simulations \#1, \#2 and \#3). 
This is expected as the film thickness is just 120\,nm and already in scenario \#2 ($\Delta w=1$\,µm), the uncovered surface is about 10$\times$ larger.
Subsequently we fitted the simulated data to Equation~\ref{eqn:FickSolution}. For scenario \#1--\#3, the origin corresponds to the beginning of the capping, $x_0$, while it is shifted towards the trench position, ${x_0'}$, or towards the inside of the sample, ${x_0''}$, for the simulations \#4 and \#5.
We determine the accuracy of the evaluation method by comparing input and fitting values for $k^*$, $D^*$ and $k_\textrm{leak}^*$, as given in the Table of Figure~\ref{fig:COMSOL}. 
The apparent (fitted) $k^*$ is strongly affected by partially re-opening the top surface and fitted values deviate from the input value up to two orders of magnitude. 
On the other hand, fitting values for $D^*$ and $k_\textrm{leak}^*$ match very well with the input values for all simulation scenarios at all temperatures. Thus, the determination of $D^*$ and $k_\textrm{leak}^*$ is robust and uncertainties and imperfections in the sample geometry have only minor impact on the extracted values. On the other hand, precautions must be taken to preserve the integrity of the capping layer and high spatial resolution is required for accurate determination of the trench position to obtain meaningful $k^*$ values. The latter can be achieved by starting the in situ line scan already a few µm inside the trench rather than just at the sample edge, at the cost of loosing time resolution.

\begin{figure*}[!hb]
	\centering
	\begin{overpic}[width=170mm, unit=1mm]
		{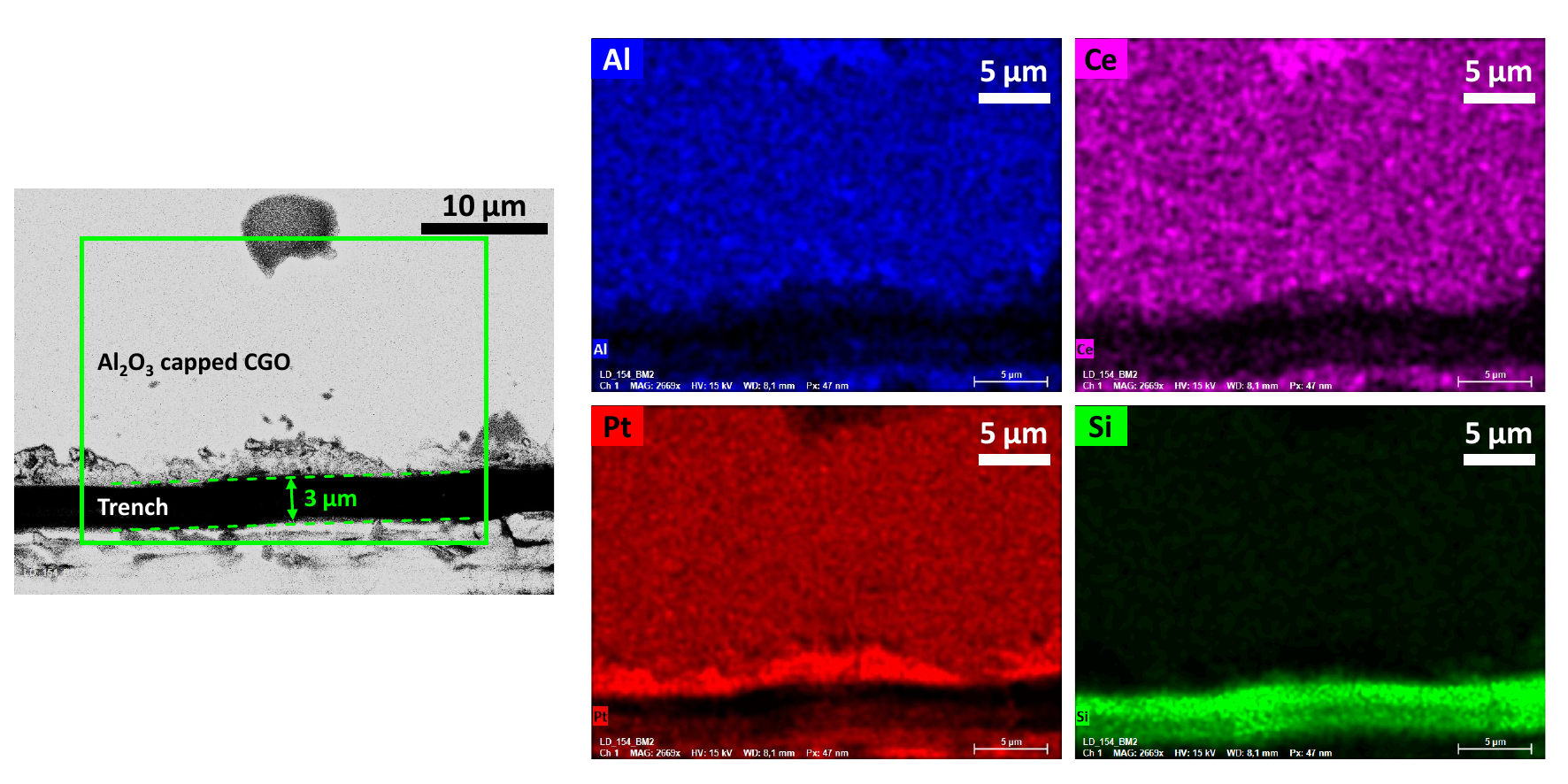}
	\end{overpic}
	\caption[]{SEM and EDX analysis of trench cut into Al$_2$2O$_3$ coated CGO thin film. The strong Si signal and the vanishing of the Al, Ce and Pt signals show that the trench cuts sufficiently deep to fully open the CGO lateral surface.}
	\label{fig:SI:SEM_trench}
\end{figure*}

While EDX top view mapping did not indicate any large areas where the signal intensity of Al and Ce did not coincide (compare Figure~\ref{fig:SI:SEM_trench}), any alteration of the capping layer close to the trench, such as chipping, could not be excluded to contribute to the observed higher $k^*$. As mentioned above, the fact that a purely pristine surface was used during the back-exchange could also add to the observation of apparent faster kinetics. 

However, we want to emphasize, that the determination of the diffusion coefficient is little affected by any of these factors, and in-plane IERS measurements allow to obtain reliable $D^*$ values in a fast, cheap and rather simple manner.

\section{Conclusions}
We have described a new in situ approach to study the in-plane self-diffusion coefficient as a function of time using isotope exchange Raman spectroscopy (IERS) in thin film samples. We report $D^*$ coefficients of Gd doped ceria thin films and validate them via comparison with conventional ex situ measurements and literature data. FEM simulations confirm the robustness of the described method, while being cheaper, faster and more versatile compared to the standard IEDP-SIMS approach. The possibility of measuring $D^*$ as a function of time is expected to advance our understanding of physicochemical processes in functional materials with the ultimate goal to improve their performance and enhance their durability.	

\section{Experimental Section} \label{section:results}
Polycrystalline 120\,nm thick Ce$_{1-x}$Gd$_{x}$O$_{2}$ ($x=0.2$) films were fabricated by large-area pulsed laser deposition (PVD Systems, PLD 5000) equipped with a 248\,nm KrF excimer laser (Lambda Physics, COMPex PRO 205) under the following conditions: temperature 700\,°C, oxygen pressure $7\cdot10^{-3}$\,mbar, target to substrate distance 90\,mm, laser fluency $\approx$1.2\,cm$^{-2}$, 10\,Hz repetition rate.
The films were deposited on top of 10$\times$10\,mm$^2$ MgO (100) polished single crystal substrates (CrysTec GmbH, Germany) and Si (111) wafers, coated with a Pt (150\,nm)/TiO$_2$ (40\,nm)/SiO$_2$ (500\,nm) multilayer (LETI, Grenoble, France) and cut into several pieces using a water cooled diamond saw. 
Conformal surface blocking layers with thicknesses between 40 and 100\,nm were deposited by atomic layer deposition (ALD) between 180 and 250\,°C and plasma enhanced chemical vapor deposition (PECVD) at 280\,°C on top of the $^{18}$O enriched CGO samples. Trenches were opened at the center of each sample with a precision diamond scriber.

The isotope exchange was performed in a tubular furnace at 640\,°C for 0.5\,h using a sealed quartz tube filled with $^{18}$O enriched oxygen gas ($c(^{18}O)=98$\,\%, $p=0.2$\,atm, CK Isotopes, UK) and fast heating (100\,°C\,min$^{-1}$) and cooling (quenching by rolling off) ramps.
If not stated otherwise, pre-equilibration steps in natural oxygen, as commonly performed for tracer diffusion experiments, were omitted, as a high $^{18}$O isotopic fraction was desired. 
For CGO, the oxygen vacancy concentration within the studied temperature window is fixed by the amount of Gd doping \cite{Yashiro2002,Wang1998}. A temperature difference between the initial exchange and the subsequent back-exchange, is therefore not expected to cause an additional chemical driving force.

The chemical and isotopic composition was obtained by SIMS analysis, which was performed using a ToF-SIMS V instrument (ION-TOF GmbH Germany) equipped with a Bi liquid metal ion gun (LMIG) for analysis and a caesium (Cs$^+$) gun for sputtering. Negative secondary ions were collected and data was obtained in burst mode operation (6 pulses), applying Poisson correction.
Charge effects were compensated by means of a 20\,eV pulsed electron flood gun. 
Depth profiling was performed by alternating sputtering of a 300$\times$300\,mm$^2$ surface area with the Cs$^+$ ion beam (2\,keV, 110\,nA) and chemical analysis with the Bi$^{3+}$ primary ion beam (25\,keV, 0.25\,pA, rasterized surface area: 50$\times$50\,µm$^2$). Generally, two SIMS profiles per sample were recorded to confirm its homogeneity.

Raman spectroscopy measurements were performed using the blue excitation wavelength (488\,nm) of a Renishaw inVia Qontor spectrometer, with a focused spot size of approximate 1\,µm$^2$. A 50$\times$ (long working distance) and a 100$\times$ objective were used for in situ studies and room temperature measurements, respectively.
The maximum laser power on the sample surface did not exceed 1\,mW. Raman measurements were performed using Renishaw's LiveTrack option to ensure sample focus at all time.
The spectrometer was calibrated at room temperature using a silicon reference sample with a theoretical position of 520.7\,cm$^{-1}$.

For in situ back-exchange measurements, a temperature cell (Linkam THMS 600 and Nextron MPS-CHH) was mounted onto the motorized Raman stage. The cells are equipped with a ceramic heater (diameter of 1" Linkam, 0.5" Nextron) and a sapphire window for Raman measurements.
Fast heating ramps between 20 and 100\,°C\,min$^{-1}$ were used to minimise any isotopic exchange during the heating, while at the same time avoid cracking of the multilayer sample.
Automatic Raman acquisition was typically started within 30-60\,s after reaching the exchange temperature, with acquisition times of 30-60\,s.
Isotopic back-exchanges were performed under flowing dry air ($<100$\,ml/min) of natural $^{18}$O composition at atmospheric pressure.

\medskip
\textbf{Supporting Information} \par 
Supporting Information is available from the Wiley Online Library or from the author.

\medskip
\textbf{Data availability} \par
All sample datasets and materials related to this work are made available under CC BY 4.0 license in the zenodo repository: 10.5281/zenodo.11028322.

\textbf{Author contributions statement} \par
Conceptualization: AS, MB; sample preparation: AS, FC, KK; investigation: AS, NN, CP; formal analysis: AS; methodology: AS; original draft: AS; review \& editing: all authors 

\medskip
\textbf{Acknowledgements} \par 
This work has received funding from the European Union's Horizon 2020 research and innovation program under grant agreement no. 824072 (Harvestore). 
We acknowledge Florence Robaut and Laetitia Rapenne for FIB lamella preparation and TEM analysis. 
This research has benefited from characterization equipment of the Grenoble INP - CMTC platform supported by the Centre of Excellence of Multifunctional Architectured Materials "CEMAM" n°ANR-10-LABX-44-01 funded by the "Investments for the Future" Program. In addition, this work has been performed with the help of the “Plateforme Technologique Amont” de Grenoble, with the financial support of the “Nanosciences aux limites de la Nanoélectronique” Fundation" and CNRS Renatech network. Chevreul institute (FR 2638), the French ministry of research, the Région Hauts de France and FEDER are acknowledged for supporting and funding the surface analysis platform (ToF-SIMS). F.C. acknowledges funding from a Marie Skłodowska Curie Actions Postdoctoral Fellowship grant (101107093).

\medskip

\bibliographystyle{MSP}
\bibliography{library}

\begin{thebibliography}{10}
\bibitem{footnote1}
\bibinfo{title}{We use the term exchange to describe the initial annealing in
  an $^{18}$o enriched atmosphere to fill the material with the tracer isotope,
  while back-exchange corresponds to a second, subsequent annealing in an
  atmosphere of natural $^{18}$o abundance to deplete the tracer within the
  sample.}
\end{thebibliography}

\beginsupplement
\clearpage
\onecolumn
\section*{Supplementary information}
\begin{Large}{Real time observation of oxygen diffusion in CGO thin films using spatially resolved Isotope Exchange Raman Spectroscopy}\end{Large}
\\
\vspace*{5mm} 

Alexander Stangl$^{1,*}$, Nicolas Nuns$^{2}$, Caroline Pirovano$^{2}$, Kosova Kreka$^{3}$, Francesco Chiabrera$^{3}$, Albert Tarancón$^{3,4}$ and Mónica Burriel$^{1,*}$
\\
\vspace*{3mm} 
$^1$Univ. Grenoble Alpes, CNRS, Grenoble-INP, LMGP, 38000 Grenoble France \\
$^2$Univ. Lille, CNRS, Centrale Lille, Univ. Artois, UMR 8181 – UCCS – Unité de Catalyse et Chimie du Solide, F-59000 Lille, France\\
$^3$Catalonia Institute for Energy Research (IREC), Barcelona, Spain\\
$^4$ICREA, 23 Passeig Lluis Companys, 08010 Barcelona, Spain\\
\vspace*{3mm} 
$^*$alexander.stangl@grenoble-inp.fr, monica.burriel@grenoble-inp.fr

\begin{figure*}[!hb]
	\centering
	\begin{overpic}[width=175mm, unit=1mm]
		{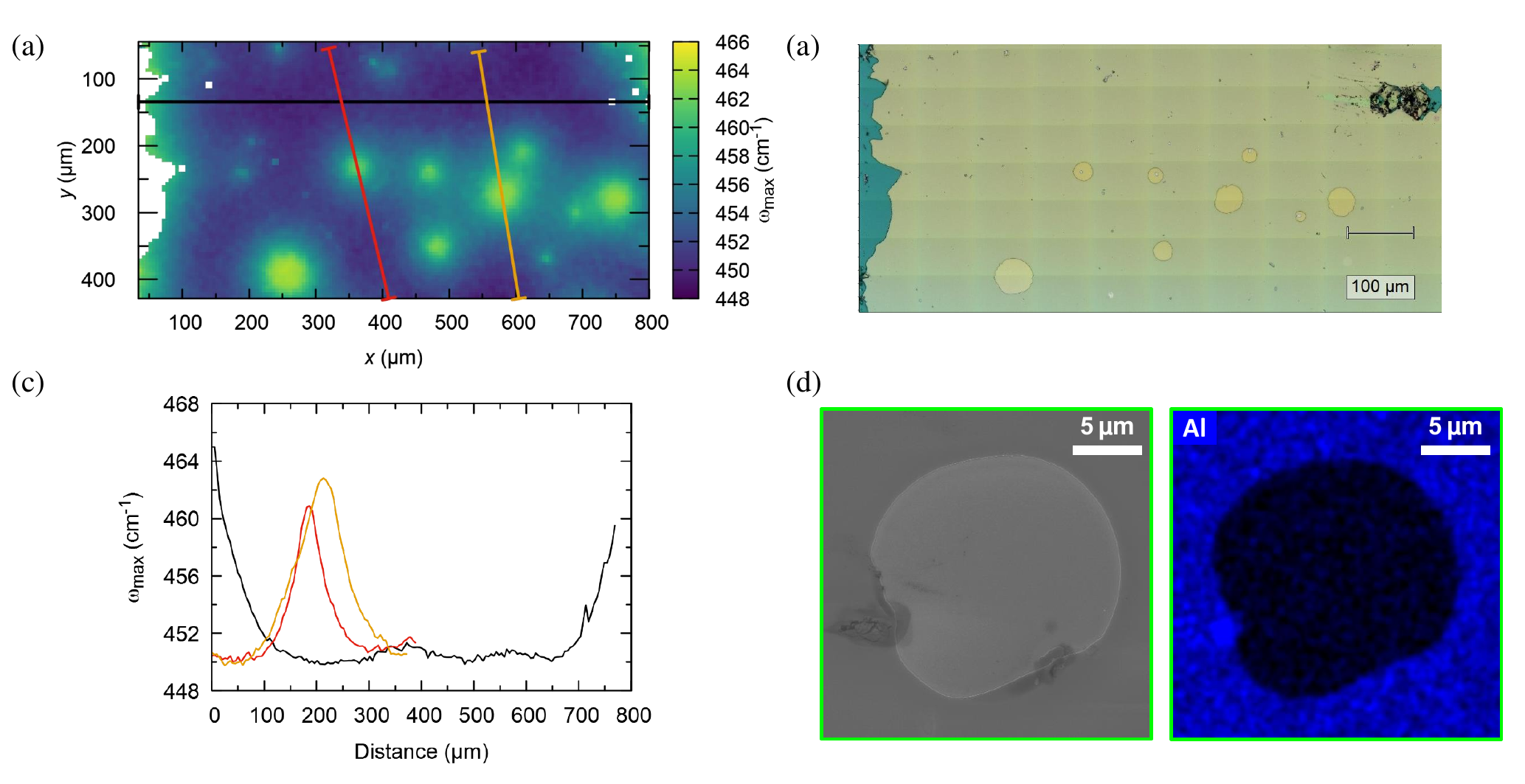}
	\end{overpic}
	\caption[]{Formation of blisters in capping layer as observed for Si$_3$N$_4$ and AlO$_2$ thin films during heating ramps above 100\,°C independent of the heating ramp. (a) Room temperature Raman map of surface after back-exchange: bright spots indicate the failure of the capping. (b) Optical microscope image of the same sample area shows clearly the same defected areas. (c) Three line profiles extracted from (a) using Gwyddion software tool. The profiles show the typical features of diffusion profile and are caused by fast exchange in the area of the blister and subsequent radial diffusion into the thin film. (d) SEM/EDX image of blister in AlO$_2$ coating. EDX image indicates that the alumina coating was completely removed due to the cracking.}
	\label{fig:SI:blister_coating}
\end{figure*}

\begin{figure*}[t]
	\centering
	\begin{minipage}[b]{0.49\textwidth}
		\begin{overpic}[width=\textwidth, unit=1mm]
			{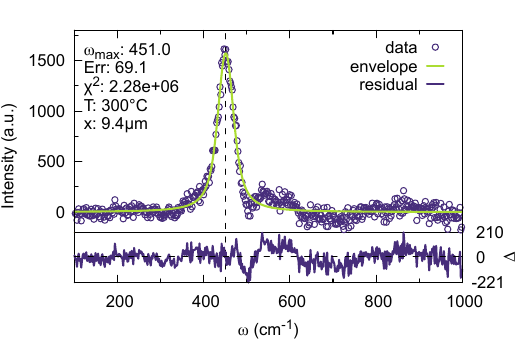}
			\put(-2.5,55){(a)}
		\end{overpic}
	\end{minipage}
	\begin{minipage}[b]{0.49\textwidth}
		\begin{overpic}[width=\textwidth, unit=1mm]
			{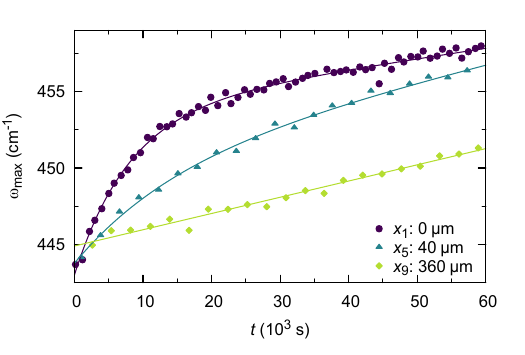}
			\put(-2.5,55){(b)}
		\end{overpic}
	\end{minipage}
	\caption[]{(a) Peak fitting of background corrected Raman spectrum to obtain mode position, $\omega_\textrm{max}(x_\textrm{i},t)$ using a Voigt profile. (b) Interpolation of $\omega_\textrm{max}(t)$ transients for three measured position $x_\textrm{i}$ using analytic function for discretization in time to increase accuracy for small $t$, where changes in $\omega_\textrm{max}$ between subsequent points is large (i.e. at small times, $t$). The interpolation is performed by fitting an analytic function (solid lines) to the data. Here, two parallel, weighted exponential decays were used, whereas the function itself and corresponding fit parameters have no physical meaning and only serve to obtain the best possible interpolation.}
	\label{fig:SI:IERS}
\end{figure*}

\end{document}